\documentclass[10pt]{article}
\usepackage{amssymb}
\usepackage{epsfig,graphicx}
\usepackage{enumerate}
\usepackage{url} 

\newcommand{\blind}{0}

\newcommand{\Prob}{{\rm I\hspace{-0.8mm}P}}
\newcommand{\Exp}{{\rm I\hspace{-0.8mm}E}}

\newcommand{\iz}{{\rm \rlap Y\kern 2.2pt Y}}

\newcommand{\RL}{{\rm I\hspace{-0.8mm}R}}

\newcommand{\bfA}{\mbox{\boldmath $A$}}

\newcommand{\bfC}{\mbox{\boldmath $C$}}
\newcommand{\bfD}{\mbox{\boldmath $D$}}

\newcommand{\bfI}{\mbox{\boldmath $I$}}
\newcommand{\bfJ}{\mbox{\boldmath $J$}}

\newcommand{\bfM}{\mbox{\boldmath $M$}}
\newcommand{\bfP}{\mbox{\boldmath $P$}}
\newcommand{\bfQ}{\mbox{\boldmath $Q$}}

\newcommand{\bfS}{\mbox{\boldmath $S$}}

\newcommand{\bfY}{\mbox{\boldmath $Y$}}

\newcommand{\e}[1]{{\rm e}^{{#1}}}
\newcommand{\refs}[1]{(\ref{#1})}

\newcommand{\proof}{\noindent {\it Proof.\ }}
\newcommand{\halmos}{\hfill $\Box$}
\newcommand{\ind}{1\hspace{-1mm}{\rm I}}

\renewcommand{\theequation}%
{{\rm
\arabic{equation}}}

\renewcommand{\baselinestretch}{1.15}

\newcounter{mylistcnt}
\renewcommand{\themylistcnt}{{\rm({\roman{mylistcnt}})}}

\newcounter{zad}

\newtheorem{Th}{Theorem}

\newtheorem{Lemma}{Lemma}

\newtheorem{Defin}{Definition}

\newtheorem{D}{}[section]

\newcommand{\cf}{{c\!f}}

\begin{document}

\def\spacingset#1{\renewcommand{\baselinestretch}%
{#1}\small\normalsize} \spacingset{1}


\if0\blind
{
  \title{\bf Premium valuation for a multiple state model \\ containing manifold premium-paid states}
  \author{Joanna D\c{e}bicka \thanks{
    Corresponding author. E-mail: joanna.debicka@ue.wroc.pl.
    This work was partially supported by The National Science Centre Poland under Grant
    2013/09/B/HS4/00490 }\\
    Department of Statistics, Wroclaw University of Economics\\
    and \\
   Beata Zmy\'{s}lona \\
    Department of Statistics, Wroclaw University of Economics}  \maketitle
} \fi

\if1\blind
{
  \bigskip
  \bigskip
  \bigskip
  \begin{center}
    {\LARGE\bf Premium valuation for a multiple state model \\ containing manifold premium-paid states}
\end{center}
  \medskip
} \fi

\bigskip
\begin{abstract}

The aim of this contribution is to derive a general matrix formula
for the net period premium paid in more than one state.
For this purpose we propose to combine actuarial technics with the
graph optimization methodology.
The obtained result is useful for example to more advanced models
of dread disease insurances allowing period premiums paid
by both healthy and ill person (e.g. not terminally yet).
As an application, we provide analysis of dread disease
insurances against the risk of lung cancer based on the actual data for the Lower
Silesian Voivodship in Poland.

\end{abstract}

\noindent%
{\it Keywords:}  modified multiple state model; Dijkstra algorithm; net premium;
stochastic interest rate; critical illness insurance; Accelerated Death Benefits.


\spacingset{1.15} 

\section{Introduction}\label{sec:intro}

The insurance market is constantly expanding.
Insurers offer more flexible contracts taking into account various situations
that may arise in life. An example would be a serious illness, in case of which,
the priorities of the insured person may change considerably.
In particular, it may be that
the death benefit becomes less important while the life benefit becomes most important.
On the insurance marked exist different kind of solutions
to protect the insured against financial problems in this difficult situation.
According to one of them, an insurer in such unexpected situations during the insurance period
may offer
purchase of an additional option called Accelerated Death Benefits (ADBs) to life insurance policyholder,
which provides an acceleration of all or of a part of the basic death benefit
to the insured before his death.
By another alternative, the insured may buy dread disease insurance (or critical illness insurance)
which provides the policyholder with a lump sum in case of dread disease which is
included in a set of diseases specified by the policy conditions,
such as heart attack, cancer or stroke (see \cite{DG93},
\cite{HP99}, \cite{Pit94}, \cite{Pit14}).
It implies that the dread disease policy does not meet any specific needs and
does not protect the policyholder against such financial losses as
loss of earnings or reimbursement of medical expenses.
In both cases conditions of this insurance products state that
the benefit is paid on diagnosis of a specified condition, rather than on disablement.
This is understandable, because this type of insurance
is sensitive to the development of medicine, not all
dread diseases are as mortal as a few years ago. Thus insurers introduce
strict conditions for the right to receive benefits associated
with a severe disease.
One of popular conditions is that benefits are paid not only on the
diagnosis but also on the disease stage that directly depends on
the expected future lifetime of a sick person. Then the
insurer has to take into account that probability of death
of a dread disease sufferer depends on
the duration of the disease.
Depending on the conditions insurance premium may be paid in various forms by:
healthy or sick (but not terminally) person, living person or healthy person.
This article focuses on accurate valuation
of such insurance products.

Multiple state modelling is a stochastic tool for
designing and implementing insurance products. The multistate
methodology is commonly used in calculation of actuarial values of
different types of life and health insurances. A general approach
to calculation of moments of the cash value of the future payment
streams (including benefits, annuities and premiums) arising from
a multistate insurance contract can be found in e.g. \cite{Deb13}.
This methodology, developed for  the discrete-time model (where
insurance payments are excercised at the end of time intervals), is
based on an {\it modified multiple state model} (or {\it extended
multiple state model}), for which matrix formulas for actuarial
values can be derived. This approach to costing contracts
not only makes calculations easier, but also enables us to
factorize the stochastic nature of the evolution of the
insured risk and the interest rate, which can be
observed in the derived formulas.

The aim of this contribution is to derive a general matrix formula for the net period premium
paid in more than one state,
which can be applied to any type of insurance being modeled by the multiple state model.
In a special case, when the insured pays the single premium in advance
or period premiums under the condition that he is healthy (or active),
the valuation of the contract may be done by the use of results derived in \cite{Deb13}.
More advanced models of dread disease insurances (as e.g. ADB's form)
allowing period premiums paid by both healthy and ill person (e.g. not terminally yet)
go beyond the scope of models analysed in \cite{Deb13} and need a different approach.
This paper focuses on the solution of this problem by the use of the graph optimization methodology to
find the shortest path between particular states of the model.
Note that the formulas obtained in \cite{Deb13}
are special cases of formula derived in this paper.

The paper is organized as follows. In Section \ref{sec:m.s.m}
we describe the modified multiple state model and its probabilistic structure.
This modification allows us to use matrix-form approach to costing insurance contract.
In Section \ref{sec:mfnp} we derive general matrix expressions for the net period premium
paid in more than one state. Section
\ref{sec:application} deals with the study of dread disease insurances against
the risk of lung cancer.
The modified multiple state model for dread disease insurances is presented in Section \ref{sec:actuarial.model}.
The probability structure of the analyzed model is built under conditions that the probability of death
for a dread disease sufferer depends on the duration of the
disease and the payment of benefits associated with a severe
disease depends both on the diagnosis and on the disease stage presented in \cite{DZ2015} (Section \ref{sec:lc-ex}).
In Section \ref{sec:ap.n.p}, the results obtained in Section \ref{sec:mfnp} are applied to
costing of different types of critical illness policies
based on the actual data for the Lower Silesian Voivodship in Poland.
Suggestions for further possible applications of obtained results are presented in Section \ref{sec:conc}.

\section{Multiple state model}\label{sec:m.s.m}

Following Haberman \& Pitacco \cite{HP99}, with a given insurance
contract we assign a {\it multiple state model}. That is, at any
time the insured risk is in one of a finite number of states
labelled by $1,2,...,N$ or simply by letters. Let ${\mathcal{S}}$
be the {\it state space}. Each state corresponds to an event which
determines the cash flows (premiums and benefits). Additionally,
by ${\mathcal{T}}$ we denote a {\it set of direct transitions}
between states of the state space.  Thus ${\mathcal{T}}$ is a
subset of the set of pairs $\left( {i,j} \right)$, i.e.,
${\mathcal{T}} \subseteq \{  \left( {i,j} \right)\mid i \neq j;
i,j \in {\mathcal{S}} \}$. The pair
$({\mathcal{S}},{\mathcal{T}})$ is called a {\it multiple state
model}, and describes all possible insured risk events as far as
its evolution is concerned (usually up to the end of insurance).
This model is structured so that it is a possibility to assign
any cash flow arising from the insurance contract to one of
the states (annuity, premiums), or the transition between them (lump sums).
That it was possible to use matrix formulas for actuarial values,
the multiple state model must be constructed so that
each cash flow must maintain its to one of the states.
Observe that for the lump sum the information that
the insured risks is in a particular state at moment $k$
is not enough to determine the benefit at time $k$,
because we need additional information about where the insured risk was at previous moment $k-1$.
Matrix is a two-dimensional structure,
thus it is not possible to determine the exact moment
of realization of lump sum benefit by using above three pieces of information.
It appears that each $({\mathcal{S}},{\mathcal{T}})$ model can be easily
(by the recursive procedure proposed in \cite{Deb13})
extended to {\it modified multiple state model} $({\mathcal{S}}^{\ast},{\mathcal{T}}^{\ast})$
in which the lump sum benefit is affiliated with particular state
and not a direct transition between states.

In this paper we consider an insurance contract issued at time $0$
(defined as the time of issue of the insurance contract) and
terminating according to the plan at a later time $n$ ($n$ is the
term of policy). Moreover, $x$ is the age of the insured person
at a policy issue.

We focus on discrete-time model. Let $X^{\ast}(k)$ denote the
state of an individual (the policy) at time $k$ ($k \in
\textrm{T}= \{0,1,2,\dots ,n\}$). Hence the evolution of the
insured risk is given by a discrete-time stochastic process $\{
{X^{\ast}(k); k\in \textrm{T}} \}$, with values in the finite set
$\mathcal{S}^{\ast}=\{ 1, 2,...,N^{\ast}\}$.
In order to describe
the probabilistic structure of
$\{X^{\ast}(k)\}$, for any moment $k \in \{0,1,2,...,n \}$, we introduce
$\Prob_{j}^{\ast}(k)=\Prob(X^{\ast}(k)=j)$ and vector
\[ \bfP(k) =(\Prob_{1}^{\ast}(k), \Prob_{2}^{\ast}(k), \Prob_{3}^{\ast}(k), \ldots ,\Prob_{N^{\ast}}^{\ast}(k))^T \in \RL^{N^{\ast}}. \]
Note that $\bfP(0) \in \RL^{N^{\ast}}$ is a vector of the initial
distribution (usually it is assumed that state $1$ is an initial
state, that is $ \bfP(0) =(1,0,0, \ldots ,0)^T \in
\RL^{N^{\ast}}$).

Under the assumption that $\{X^{\ast}(k)\}$ is a
nonhomogeneous Markov chain (see, e.g. \cite{Deb13}, \cite{Hoe69}, \cite{Hoe88}, \cite{Wat84},
\cite{Wol94}) we have $\bfP^T(t)=\bfP^T(0)\prod^{t-1}_{k=0}\bfQ^{\ast}(k),$
where $\bfQ^{\ast}(k)=(q_{ij}^{\ast}(k))_{i,j=1}^{N^{\ast}}$ with
$q_{ij}^{\ast}(k)=\Prob (X^{\ast}(k+1)=j | X^{\ast}(k)=i)$ being the
transition probability.
The above transition probabilities can be determined
using a {\it multiple increment-decrement table} (or {\it
multiple state life table}).

\section{Matrix formula for net premiums}\label{sec:mfnp}

Before presenting the matrix formula for the net period premium we
need to introduce some notation (cf. \cite{Deb13}).

In order to describe the probabilistic structure of $\{X^{\ast}(k)\}$
we introduce matrix
\begin{eqnarray}\label{D}
\bfD= \left(
 \begin{array}{l}
  \bfP(0)^T \\
  \bfP(1)^T \\
 \multicolumn{1}{c}{\dotfill}\\
  \bfP(n)^T
 \end{array}
\right) \in \RL^{(n+1) \times (N^{\ast})}.
\end{eqnarray}.

The individual's presence in a given state may have some financial
effect. For $k$-th unit of time (it means for period $[k-1,k)$), we
distinguish between the following types of cash flows: a cash flow
paid in advance at time $k-1$ if $X^{\ast}(k-1)=i$ (premiums and
life annuity due) and a cash flow paid from below at time $k$ if
$X^{\ast}(k)=j$ (lump sum and immediate life annuity). Note that
the insurance policy gives rise to two payment streams. Firstly, a
stream of premium payments, which flows from the insured to the
insurer. Secondly, in the opposite direction, a stream of actuarial
payment functions, where fixed amounts under the annuity product
and lump sum benefits are considered as a series of deterministic
future cash flows.

One of important quantities is
the {\it total loss} ${\mathcal{L}}$ of the insurance contract,
defined as the difference between the present value of future
benefits and the present value of future premiums. In particular,
the stream of actuarial payment functions is an {\it inflow}
representing an income to ${\mathcal{L}}$ and it takes positive
values, while the stream of premium payments is an {\it outflow}
representing an outgo from ${\mathcal{L}}$ and it takes negative
values.

Let ${\cf}_{\! \! j}^{\ast}(k)$ be the future cash flow payable at
time $k$  if $X^{\ast}(k)=j$ ($k= 0,1,...,n$) and
\begin{eqnarray*}
\bfC= \left(
 \begin{array}{llcl}
  {\cf}_{1}^{\ast}(0) & {\cf}_{2}^{\ast}(0) & \cdots & {\cf}_{N^{\ast}}^{\ast}(0)\\
  {\cf}_{1}^{\ast}(1) & {\cf}_{2}^{\ast}(1) & \cdots & {\cf}_{N^{\ast}}^{\ast}(1)\\
  {\cf}_{1}^{\ast}(2) & {\cf}_{2}^{\ast}(2) & \cdots & {\cf}_{N^{\ast}}^{\ast}(2)\\
  \multicolumn{4}{c}{\dotfill}\\
  {\cf}_{1}^{\ast}(n) & {\cf}_{2}^{\ast}(n) & \cdots & {\cf}_{N^{\ast}}^{\ast}(n)
 \end{array}
\right)
\end{eqnarray*}
denote $(n+1)\times N^{\ast}$ cash flows matrix.

From the financial point of view, the cash flow ${\cf}_{\!
\!i}^{\ast}(k)$ is a sum of {\it inflows} representing an income
to a particular fund and {\it outflows} representing an outgo from
a particular fund. Hence
\begin{eqnarray}\label{C.in.out}
\bfC=\bfC_{in}+\bfC_{out},
\end{eqnarray}
where $\bfC_{in}$ consists only of an income to a particular fund
and $\bfC_{out}$ consists only of an outgo from a particular fund.
We note that for ${\mathcal{L}}$, $\bfC_{in}$ includes the
benefits and $\bfC_{out}$ includes the premiums.

Let $Y(t)$ denote the rate of interest in time interval $[0,t]$.
Then the discount function $v(t)$ is of the form
$\upsilon(t)=\e{-Y(t)}$. It is useful to introduce the following notation
\[ \bfY=(e^{-Y(0)},e^{-Y(1)} , ... ,e^{-Y(n)})^T \in \RL^{n+1} \]
and
\[ \Exp (Y) = \bfM = (m_0, m_1, ... ,m_n)^T \in \RL^{n+1}, \]
with $m_k=\Exp(e^{-Y(k)})$. We refer to \cite{Deb03} for the exact
forms of the matrix $\bfM$ when $Y(t)$ is modelled by an
Ornstein-Uhlenbeck process or a Wiener process. Let us note that
for constant interest rate, we have $\upsilon(0,k)=\upsilon^k$ and
$\bfM=(1,\upsilon,\upsilon^2,...,\upsilon^n)^T$.

Additionally, let
\begin{eqnarray*}
\bfS &=&(1, 1, 1, \ldots ,1)^T \in \RL^{N^{\ast}}, \\
\bfI_{k+1} &=&(0, 0, \ldots, \underbrace{1}_{k+1}, \ldots ,0)^T \in \RL^{n+1}, \\
\bfJ_j &=&(0, 0, \ldots, \underbrace{1}_j, \ldots ,0)^T \in
\RL^{N^{\ast}},
\end{eqnarray*}
for each $j=1,2,..., N^{\ast}$ and $k=0,1,2,...,n$.

Furthermore, for any matrix $\bfA
=\left(a_{ij}\right)_{i,j=1}^{n+1}$ let $\emph{Diag}(\bfA)$ be a
diagonal matrix
\begin{eqnarray*}
 \emph{Diag}(\bfA) =
\left(
 \begin{array}{cccc}
   a_{11} & 0      & \cdots & 0 \\
   0      & a_{22} & \ldots & 0 \\
   \multicolumn{4}{c}{\dotfill}\\
   0      & 0      & \ldots & a_{n+1n+1} \
 \end{array}
\right).
\end{eqnarray*}

Insurance premiums are called {\it net premiums} if the {\it
equivalence principle} is satisfied, i.e. $\Exp ({\mathcal{L}})=
0$. In order to study the first moment of ${\mathcal{L}}$ we make the
following standard assumptions (see also \cite{Deb13},
\cite{Fre90} or \cite{Par94}):

\begin{description}
\item[{\bf Assumption A1}] Random variable $X^{\ast}(t)$ is independent of $Y(t)$.
\item[{\bf Assumption A2}] First moment of the random discounting function $e^{-Y(t)}$
                is finite.
\end{description}

The {\it net single premium} paid in advance (at time $0$, when
$X^{\ast}(0)=1$) for the insurance modelled by $(\mathcal{S}^{\ast},\mathcal{T}^{\ast})$
equals (cf. \cite{Deb13})
\begin{eqnarray}\label{sk³.jednoraz.}
\pi = \pi_1(0) = \bfM^T Diag \left(\bfC_{in} \bfD^T \right)\bfS.
\end{eqnarray}

Additionally, by \cite{Deb13}, the {\it net period premium} payable in advance at the beginning
of the time unit during the first $m$ units ($m \leq n$) if
$X^{\ast}(t)=1$ equals
\begin{eqnarray}\label{skladkap}
p=\frac{\bfM^T Diag \left(\bfC_{in} \bfD^T \right)\bfS}{ \bfM^T
\left[ \bfI - \sum_{k=m+1}^{n+1} \bfI_k \bfI_k^T \right] \bfD
\bfJ_1},
\end{eqnarray}
where the denominator in \refs{skladkap} is equal to the actuarial
value of a temporary ($m$-year) life annuity-due contract
$\ddot{a}_{11}(0,m-1)$ .

In case of each type of insurance, the net single premium can be calculated
using formula \refs{sk³.jednoraz.}. Importantly, formula for net
period premium has to be modified, because premiums may be paid
not only if $X^{\ast}(k)=1$, but also
when $X^{\ast}(k)$ is in other states
(of course those in which an insured person is alive). We derive
formula for period premium in Theorem \ref{Tw1}.

Let $\ddot{a}_{1(i)}(k_1,k_2)$ denote the actuarial value of the
stream of unit benefits arising from  life annuity-due contract
payable in period $[k_1,k_2)$ if $X^{\ast}(k)=i$ for $k \in
[k_1,k_2)$. Actuarial value is calculated at the beginning of the
insurance period ($k=0$). We tacitly assume that $X^{\ast}(0)=1$.

\begin{Lemma}\label{Lemma1}

Suppose that A1-A2 hold and $X^{\ast}(0)=1$. Then for $(\mathcal{S}^{\ast},\mathcal{T}^{\ast})$ we
have
\begin{eqnarray}\label{a-1i}
\ddot{a}_{1(i)}(k_1,k_2)=\bfM^T \left( \sum_{t=k_1}^{k_2-1}
\bfI_{t+1} \bfI_{t+1}^T \right) \bfD \bfJ_i.
\end{eqnarray}
\end{Lemma}

\proof Let us observe that under assumption A1-A2 we have
\begin{eqnarray}\label{a-1i-1}
\ddot{a}_{1(i)}(k_1,k_2) =\sum_{t=k_1}^{k_2-1} \Exp \left(
\e{-Y(t)}\right) \cdot \Prob(X^{\ast}(t)=i \mid X^{\ast}(0)=1).
\end{eqnarray}
Since $X^{\ast}(0)=1$, then $\Prob(X^{\ast}(t)=i \mid X^{\ast}(0)=1)=
\Prob_{j}^{\ast}(t)$. Moreover
\begin{eqnarray}
\Exp \left(  \e{-Y(t)}\right)&=& \bfM^T \bfI_{t+1}, \label{v}\\
\Prob_{i}^{\ast}(t)          &=& \bfI^{T}_{t+1} \bfD \bfJ_{i}
\label{p}.
\end{eqnarray}
Applying \refs{v} and \refs{p} to \refs{a-1i-1} we have
\begin{eqnarray}
\ddot{a}_{1(i)}(k_1,k_2) &=& \sum_{t=k_1}^{k_2-1} \bfM^T
\bfI_{t+1}\bfI^{T}_{t+1} \bfD \bfJ_{i} = \bfM^T
\left(\sum_{t=k_1}^{k_2-1} \bfI_{t+1}\bfI^{T}_{t+1} \right) \bfD
\bfJ_{i}, \nonumber
\end{eqnarray}
which completes the proof of \refs{a-1i}.

\halmos

In order to determine the value of premium payable when the process
$\{X^{\ast}(k)\}$ is at state $i$ in time interval $[k_1,k_2)$, it
is necessary to designate the shortest possible sequence of
transitions from state $1$ to state $i$ at time period $[0, k_2)$.
Note that  multiple state model reminds a directed graph, where
the states correspond to the vertices (the nodes) of the graph,
and the direct transitions correspond to the edges between the
nodes. Therefore, in order to find the shortest way (path) between
the states we use the graph optimization methodology.

Let $w=\left((i_0,i_1),(i_1,i_2),...,(i_{k-1},i_k) \right)$, where $i_0,
i_1,...i_k \in \mathcal{S}^{\ast}$, denote a path (way) from state $i_0$ to state
$i_k$ in the model $(\mathcal{S}^{\ast},\mathcal{T}^{\ast})$.
By $d(w)$ let us denote the path length of $w$ i.e.
\begin{eqnarray}\label{d(w)}
d(w)=\sum_{(i,j)\in \mathcal{T}^{\ast}} \ind_{\{(i,j)\subset w\}}.
\end{eqnarray}
Additionally, let $\delta(i_0,i_k)=\min_{w} d(w)$ be the length of the shortest path from
$i_0$ to $i_k$, where the minimum runs throughout all the paths leading from
$i_0$ to $i_k$. Observe that the shortest path, if exists, must be
a straight path, i.e. such that all the nodes of the path are
different. This shortest path can be determined by Dijkstra's
algorithm \cite{Dij59}.

Let  $\mathcal{S}^{p}\subset \mathcal{S}^{\ast}$ be such that
$X^{\ast}(k)\in\mathcal{S}^{p}$ for $k=0,1,...,m-1$ implies that
the period premium $p_{\mathcal{S}^{p}}$ is paid. The formula for
such a premium is presented in Theorem \ref{Tw1}.

\begin{Th}\label{Tw1}
Suppose that equivalence principle holds and assumptions A1-A2 are
satisfied. Moreover, for extended multiple state model
$({\mathcal{S}^{\ast}},{\mathcal{T}^{\ast}})$ the cash flows
matrix is defined for the insurer's total loss fund, and insurance
premiums are paid for $k=0,1,2,\ldots,m-1$ if $X^{\ast}(k)=i$ and
$i \in \mathcal{S}^{p}$. Then the formula for net period premium
$p_{\mathcal{S}^{p}}$ paid during the first $m$ units of the
insurance period has the following form
\begin{eqnarray}\label{skladkap.ws}
p_{\mathcal{S}^{p}}=\frac{\bfM^T Diag \left(\bfC_{in} \bfD^T
\right)\bfS}{ \sum_{i \in \mathcal{S}^{p} \wedge \delta(1,i) <m}
\ddot{a}_{1(i)}(\delta(1,i),m)},
\end{eqnarray}
where $\ddot{a}_{1(i)}(\delta(1,i),m)=\bfM^T \left(
\sum_{k=\delta(1,i)}^{m-1} \bfI_{k+1} \bfI_{k+1}^T \right) \bfD
\bfJ_i$ and $\delta(1,i)$ is the shortest path from state $1$ to
state $i$ in $({\mathcal{S}^{\ast}},{\mathcal{T}^{\ast}})$.
\end{Th}

\proof For the multistate insurance, the equivalence
principle $\Exp
({\mathcal{L}})= 0$ may be written in the following form (Theorem 2 in
\cite{Deb13})
\begin{eqnarray}\label{zas.row.L}
\bfM^T Diag \left(\bfC \bfD^T \right)\bfS = 0, \nonumber
\end{eqnarray}
which combined with \refs{C.in.out} gives
\begin{eqnarray}\label{zas.row.C}
\bfM^T Diag \left(-\bfC_{out} \bfD^T \right)\bfS = \bfM^T Diag
\left(\bfC_{in} \bfD^T \right)\bfS.
\end{eqnarray}
Let $p=p_{\mathcal{S}^{p}}$ be the net period premium payable in
advance at the beginning of the unit time during first $m$ units,
when $X^{\ast}(k)=i$ and $i \in {\mathcal{S}^{p}}$. Since
$X^{\ast}(0)=1$, then the premium may be paid for the first time
at moment $k=\sigma(1,i)$, provided that $\sigma(1,i)<m$.
Let $\bfC_{out(p,i,m)}=-p \cdot \bfC_{(\ddot{1},i,m)}$, where matrix
$\bfC_{(\ddot{1},i,m)}$ consists of the following columns
($j=1,2,...,N^{\ast}$)
\begin{eqnarray*}
\bfC_{(\ddot{1},i,m)}\bfJ_j= \left\{
\begin{array}{lcl}
  (0,\ldots,0,\underbrace{1}_{\delta(1,i)},1,\ldots,1,\underbrace{1}_{m-1},0, \ldots,0)^T & \textrm{for} & j = i \\
  (0,\ldots,0)^T & \textrm{for} & j \neq i
\end{array}
\right. .
\end{eqnarray*}
Assuming that the premiums may be paid during the first $m$ units
of the insurance period if $X^{\ast}(k)=i$ and $i \in \mathcal{S}^{p}$, we have
\begin{eqnarray}\label{C_out-i}
\bfC_{out}=\sum_{i \in \mathcal{S}^{p} \wedge \delta(1,i)
<m}\bfC_{out(p,i,m)}= -p \sum_{i \in \mathcal{S}^{p} \wedge
\delta(1,i) <m} \bfC_{(\ddot{1},i,m)}.
\end{eqnarray}
Applying \refs{C_out-i} to left side of equation \refs{zas.row.C}
we obtain
\begin{eqnarray}
& & \bfM^T Diag \left(\left(p \sum_{i \in \mathcal{S}^{p} \wedge
\delta(1,i) <m} \bfC_{(\ddot{1},i,m)} \right)\bfD^T \right)\bfS =
\nonumber \\
& & = p \sum_{i \in \mathcal{S}^{p} \wedge \delta(1,i) <m} \bfM^T
Diag \left(\bfC_{(\ddot{1},i,m)}\bfD^T \right)\bfS \nonumber \\
& & = p \sum_{i \in \mathcal{S}^{p} \wedge \delta(1,i) <m} \bfM^T
\sum^{n}_{k=0} \bfI_{k+1} \bfI^{T}_{k+1}
\bfC_{(\ddot{1},i,m)}\bfD^T \bfI_{k+1}.\label{zas.row.C-1}
\end{eqnarray}

Note that
\begin{eqnarray}\label{zas.row.C-2}
\bfI^{T}_{k+1} \bfC_{(\ddot{1},i,m)}\bfD^T \bfI_{k+1} = \left\{
\begin{array}{lcl}
  \Prob^{\ast}_{i}(k) & \textrm{for} &
  k=\delta(1,i),\delta(1,i)+1,\ldots,m-1 \\
  0 & \textrm{for} &k =0,1,\ldots,\delta(1,i)-1 \ \textrm{and}\\
    &              &k=m,m+1,\ldots,n
\end{array}
\right. .
\end{eqnarray}
Combination of \refs{p} and \refs{zas.row.C-2} to
\refs{zas.row.C-1} leads to
\begin{eqnarray}
& & p \sum_{i \in \mathcal{S}^{p} \wedge \delta(1,i) <m} \bfM^T
\sum^{n}_{k=0} \bfI_{k+1} \bfI^{T}_{k+1}
\bfC_{(\ddot{1},i,m)}\bfD^T
\bfI_{k+1} = \nonumber \\
&& = p \sum_{i \in \mathcal{S}^{p} \wedge \delta(1,i) <m} \bfM^T
\sum^{m-1}_{k=\delta(1,i)} \bfI_{k+1} \bfI^{T}_{k+1} \bfD
\bfJ_{i}\nonumber \\
&& = p \sum_{i \in \mathcal{S}^{p} \wedge \delta(1,i) <m} \bfM^T
\left( \sum^{m-1}_{k=\delta(1,i)} \bfI_{k+1} \bfI^{T}_{k+1}
\right) \bfD \bfJ_{i}. \nonumber 
\end{eqnarray}

Then by Lemma~\ref{Lemma1} and \refs{zas.row.C} we
straightforwardly obtain \refs{skladkap.ws}. This completes the
proof.

\halmos

Theorem~\ref{Tw1} extends findings of \cite{Deb13}
to the case where premiums are paid not only in the initial state.
The matrix form derived in Theorem~\ref{Tw1} not only provides a concise formula
for $p_{\mathcal{S}^{p}}$, but also factorizes the double
stochastic nature of actuarial values of the total payment stream
arising from the insurance contract. Matrix $\bfD$ depends only on the
distribution of process $\{X^{\ast}(t)\}$, while $\bfM$ depends
only on the interest rate. Moreover, matrix $\bfC$ depends on cash
flows and describes the type (the case) of the insurance contract.

\section{Applications}\label{sec:application}

In this section we apply results derived in Theorem~\ref{Tw1} to
dread disease insurances on the example of the critical insurance against a lung cancer.

\subsection{Actuarial model for dread disease insurance}\label{sec:actuarial.model}

Dread disease (or 'critical illness') policies provide the
policyholder with a lump sum in case of dread disease which is
included in a set of diseases specified by the policy conditions,
such as heart attack, cancer or stroke (see \cite{DG93},
\cite{HP99}, \cite{Pit94}, \cite{Pit14}).
Typically conditions of this insurance products state that
the benefit is paid on diagnosis of a specified condition, rather than on disablement.
It implies that dread disease policy does not meet any specific needs and
does not protect the policyholder against such financial losses as
loss of earnings or reimbursement of medical expenses.
Individual critical illness insurance can take one of two main forms: a {\it
stand-alone} cover or a {\it rider benefit} for a basic life
insurance. The rider benefit (also called {\it living benefit}) may
provide an acceleration of all or of a part of the basic life cover
(Accelerated Death Benefits - ADBs), or it may be an additional
benefit.

\begin{figure}[bht]
\vspace{0.0cm}
\hspace{0.1cm}
\includegraphics[width=5.4in, height=3.5in]
{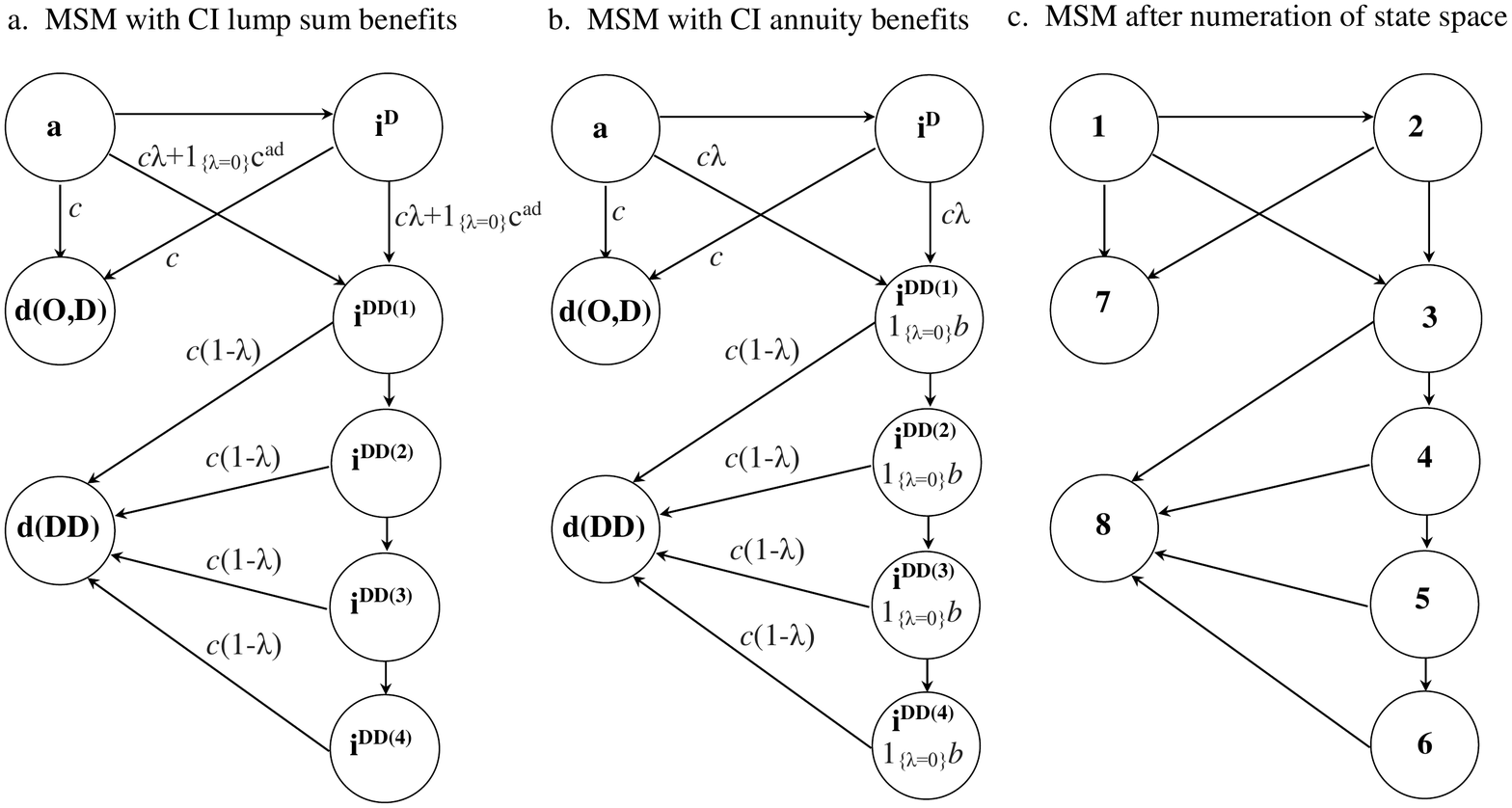}
\caption{\label{Fig.1} A multiple state model for dread disease
insurance with a DD benefit.}
\end{figure}

Dread disease insurances are of a long-term type, hence they are sensitive to the development of medicine, not all
dread diseases are as mortal as a few years ago. Thus insurers introduce
strict conditions for the right to receive benefits associated
with a severe disease. One of popular conditions is that benefits are paid not only on the
diagnosis but also on the disease stage that directly depends on
the expected future lifetime of a sick person. Then the
insurer has to take into account that probability of death of a dread disease sufferer depends on
the duration of the disease.
Let us recall that in the classical notation, used for critical illness insurances, statuses are labeled by letters,
where $a$ means that the insured is active (or healthy), $i$
indicates that the insured person is ill and suffers from a dread
disease and $d$ is related to the death of the insured; see e.g. \cite{HP99}, \cite{Pit14}.
In this paper we distinguish between states
\begin{description}
\item[$d(O,D)$ -] the death of the insured person who is ill and his
expected future lifetime is at least 4 years ($e_s\geq4$) or due to other cases, and
\item[$d(DD)$ -] the death of the insured
person who is ill and his expected future lifetime is less than
4~years ($e_s<4$),
\end{description}
where $e_s$ is the expected future lifetime of $s$-years-old person.
Moreover, following \cite{DZ2015}, state $i$ is divided into five states:
\begin{description}
\item[$i^D$ -] the insured person is ill and his
expected future lifetime is at least 4 years ($e_s\geq4$). In this
stage the remission of the disease is still possible, although
return to health state is impossible.
\item[$i^{DD(h)}$ $(h=1,2,3,4)$ -] the
insured is terminally sick and his expected lifetime is less
than $4-(h-1)$ years . In this stages the remission of the disease is
very unlikely.
\end{description}
This leads to a multiple state model for dread disease insurance derived in \cite{DZ2015}; see Figure~\ref{Fig.1}.
Note that states $i^{DD(h)}$ are reflex (that is strictly transitional and
after one unit of time, the insured risk leaves this state).
Unbundling of the four states $i^{DD(h)}$ results from the fact that
typically an insurer pays the benefit to a insured sick
whose expected future lifetime is no longer than four years or, in
some cases, two years (depending on medical
circumstances).
The difference results from the definition of a terminally ill person.
On the one hand, for example the HIV+ patients
with more than 4,5 years of life expectancy, are
treated as patients in a relatively good health.
On the other hand, the term {\it terminally ill} in
the context of health care refers to a person who is suffering
from a serious illness and whose life is not expected to go beyond
2 years at the maximum.

The general multiple state model for critical illness insurances
covers both disease lump sum benefits (Figure~\ref{Fig.1}a) and
disease annuity benefits (Figure~\ref{Fig.1}b). Next to the arcs
are marked benefits related to the transition between states,
where $c$ is a given lump sum ({\it death benefit}) and $c^{ad}$ is
an additional lump sum ({\it disease benefit}).
To avoid a situation of 'overpayment' (that
could take place when death occurs within a very short period
after disease inception to the terminal phases of the dread
disease), the single cash payment $c^{ad}$ is replaced with a
series of payments  $b$ (the annuity),
which are connected with staying of the insured risk in states $i^{DD(h)}$.
In particular,
the model presented in Figure~\ref{Fig.1}b may be applied to critical illness insurance contract
with  increasing ($b_3<b_4<b_5<b_6$) or decreasing ($b_3>b_4>b_5>b_6$) annuity benefits,
where $b_j$ is an annuity rate realized at state $j=3,4,5,6$.

By $\lambda \in [0,1]$ we denote the so called {\it acceleration parameter}.
The amount $c\lambda+\ind_{\{\lambda=0\}} c^{ad}$ is payable
after the dread disease diagnosis, while the remaining amount
$c(1-\lambda)$ is payable after death, if both random events
occur within the policy term $n$.
Note that the multiple state model presented in Figure~\ref{Fig.1} covers all
forms of DD insurances.
Namely, if $\lambda=0$, then the model describes a rider benefit as an additional benefit.
If $0< \lambda <1$, then the model describes a rider benefit as an acceleration of part of
the basic life cover.
For $\lambda=1$, the model becomes a stand-alone cover. In this case
state $i^{DD(1)}$  is absorbing, because the
whole insurance cover ceases immediately after the terminal stage dread disease
diagnosis.

In order to simplify notation, let us label states according to Figure~\ref{Fig.1}c.

\begin{figure}[h!bt]
\includegraphics[width=6in,height=4.0in]{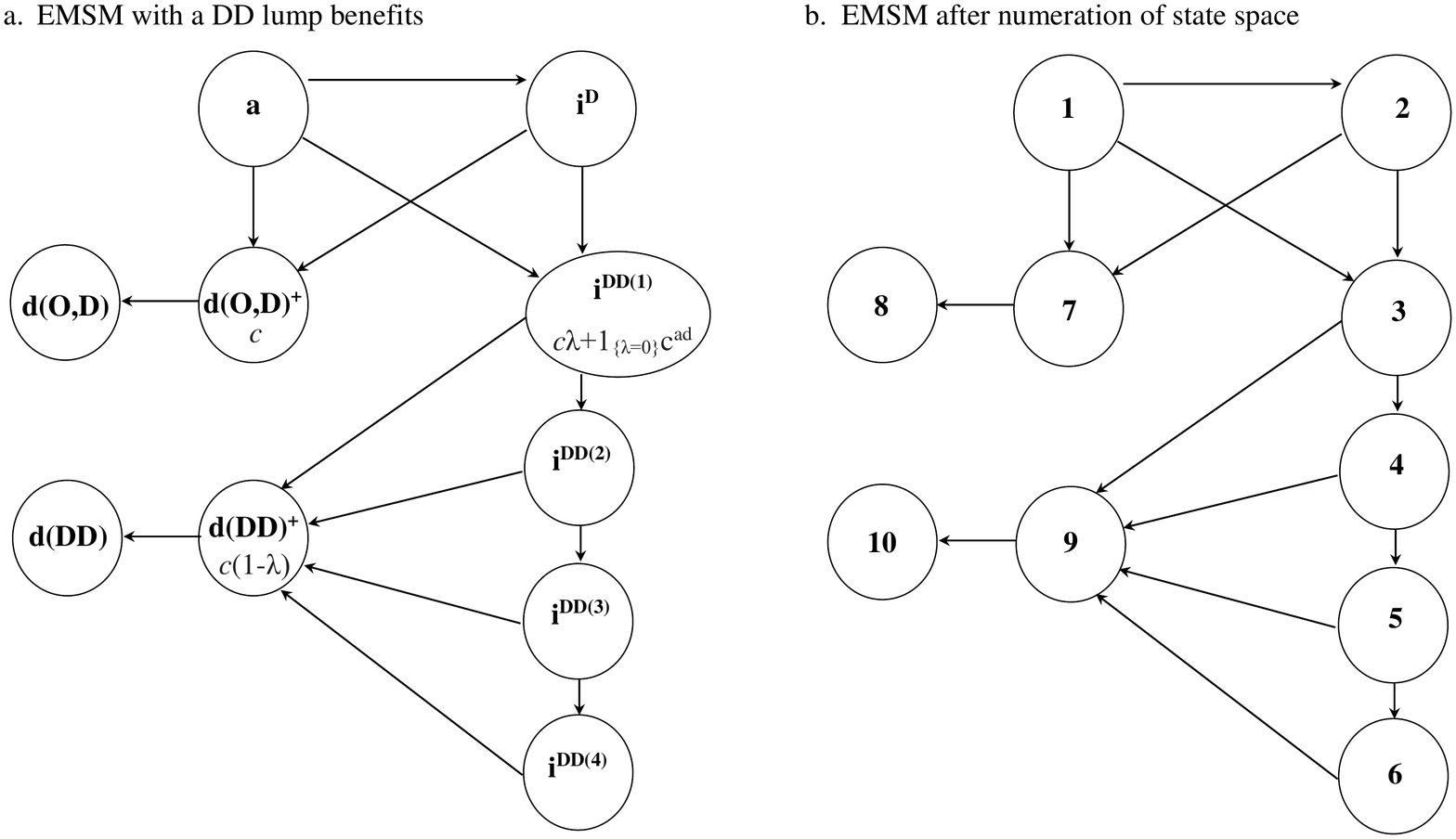}
\caption{\label{Fig.2} An extended multiple state model for dread
disease insurance.}
\end{figure}

Since the multiple state model presented in Figure~\ref{Fig.1}c is
extensive, in order to determine  appropriate actuarial values, it is worth using matrix notation.
For this purpose, we have to modify the model,
replacing lump sum benefits by benefits associated with staying of the insured risk in
particular states (according procedure presented in \cite{Deb13}).
As a result, {\it the modified
multiple state model} $(\mathcal{S}^{\ast},\mathcal{T}^{\ast})$
for dread disease insurance assumes the form presented in Figure~\ref{Fig.2}a (with DD lump sum benefits).
Following the procedure of extending the multiple
state model presented in \cite{Deb13}, we introduce states
$d(O,D)^{+}$ and $d(DD)^{+}$. State $d(O,D)^{+}$ denotes death of the insured person who is ill and his
expected future lifetime is at least 4 years ($e_s\geq4$).
If the insured risk is in this state, the death benefit $c$ is paid.
State $d(O,D)$ in this model denotes that the insured has been dead for
at least one year. Although states $d(O,D)^{+}$ and $d(D,O)$
deal with the same event,
they differ by the fact that the lump sum is realized only when the insured risk is at state $d(O,D)^{+}$. States $d(DD)^{+}$ and $d(DD)$ are interpreted
correspondingly. Note that $d(O,D)^{+}$ and $d(DD)^{+}$ are
reflex. Because $i^{DD(1)}$ is a reflex state, there is no
need to create state $i^{DD(1)^+}$.

The extended multiple state model with dread disease annuity benefits is the same as in Figure~\ref{Fig.2}a.

In order to simplify the notation in what follows we enumerate the set space, as presented in
Figure~\ref{Fig.2}b.

\subsection{Dread disease insurance against risk of lung cancer}\label{sec:lc-ex}

Malignant tumours constitute the second
cause (after cardiovascular diseases) of death in developed countries;
see \cite{DZ2015}.
In particular, lung cancer
falls into the group of tumours characterized by the highest
morbidity and mortality rates. In many European countries, it is the most frequent
in population of men and the second frequent in population of
women after breast cancer. Moreover, lung cancer is  a
tumour with unfavourable prognosis. Because of the high prevalence and
mortality rates, the relatively short survival time after the
diagnosis, lung cancer is a perfect example of the deadly disease,
which could be covered by DD insurances.

Age, sex and region of
residence should be taken into account in the analysis of the
etiology of lung cancer. Epidemiological data shows that the
morbidity rate is several times higher in male
population than in female.
Also, the incidence rate strongly depends on age. Lung cancer occurs very
rarely among  patients up to forty years of age, and then the incidence
begins to increase after the age of fifty. The peak incidence
occurs at the sixth and seventh decades of life.
The analysis of geographic data shows a significant diversity of incidence and
mortality rates to be observed in different regions of Europe. For example, in
Poland, the morbidity and mortality vary significantly among
particular provinces (voivodships).

In case of DD disease insurance for lung cancer, the model
(presented in Figure~\ref{Fig.2}b) has six states associated with
health situation of the insured person, which means that the insured:
\begin{description}
\item[$1$] - is alive and not sick with malignant lung tumour,
\item[$2$] - is diagnosed of lung cancer without
finding of metastasis to lymph nodes, brain, bones or so-called
distant metastases,
\item[$3$] - is diagnosed of lung cancer and the existence of
distant metastases are observed and his/her expected lifetime is
less than $4$ years ($e_y < 4$),
\item[$4$] - has a lung cancer with distant metastases and $e_y < 3$,
\item[$5$] - has a lung cancer with distant metastases and $e_y < 2$,
\item[$6$] - has a lung cancer with distant metastases and $e_y < 1$,
\end{description}
Other states are associated with the death of the insured person.

Following \cite{DZ2015}, in order to estimate elements of transition matrix
$\bfQ^{\ast}(k)$ we used databases  \cite{TTZ-GUS}, \cite{WD14} and \cite{NHF}.

The transition probabilities in $\bfQ^{\ast}(k)$ can be determined
using a {\it multiple increment-decrement table} (or {\it
multiple state life table}). Such a table, referring to $x$ years old
person for $({\mathcal{S}},{\mathcal{T}})$
presented in Figure~\ref{Fig.1}c, takes the following form (cf. \cite{DZ2015})
\begin{eqnarray}
\left\{l_{x+k}^{1},l_{x+k}^{2},l_{x+k}^{3},l_{x+k}^{4},l_{x+k}^{5},l_{x+k}^{6},
d_{x+k}^{12}, d_{x+k}^{13}, d_{x+k}^{17}, d_{x+k}^{23},
d_{x+k}^{27}, \right. \nonumber \\
\left. d_{x+k}^{34}, d_{x+k}^{38}, d_{x+k}^{45}, d_{x+k}^{48},
d_{x+k}^{56}, d_{x+k}^{58} \right\}_{k \geq 0} \label{mi-dlt-stare}
\end{eqnarray}
where $l^{i}_{x+k}$ denotes the number of lives in state $i$ at age $x + k$ and
$d^{ij}_{x+k}$ denotes the number of lives at
age $x+k$, who during period $[x+k, x+k+1)$ left the state $i$ and
transit to state $j$.

Notice that after application of the procedure of extension of $({\mathcal{S}},{\mathcal{T}})$,
state 8 became state 9 in $({\mathcal{S}^{\ast}},{\mathcal{T}^{\ast}})$ (cf. Figure~\ref{Fig.2}b).
Thus for $(\mathcal{S}^{\ast},\mathcal{T}^{\ast})$ the multiple increment-decrement table is as follows
\begin{eqnarray}
\left\{l_{x+k}^{1},l_{x+k}^{2},l_{x+k}^{3},l_{x+k}^{4},l_{x+k}^{5},l_{x+k}^{6},
d_{x+k}^{12}, d_{x+k}^{13}, d_{x+k}^{17}, d_{x+k}^{23},
d_{x+k}^{27}, \right. \nonumber \\
\left. d_{x+k}^{34}, d_{x+k}^{39}, d_{x+k}^{45}, d_{x+k}^{49},
d_{x+k}^{56}, d_{x+k}^{59} \right\}_{k \geq 0} \label{mi-dlt}.
\end{eqnarray}

Generally, the multiple
increment-decrement table, that refers to $x$ years old person for a
multiple state model $(\mathcal{S}, \mathcal{T})$ consists of
functions described for each transient state $i \in \mathcal{S}$.
It appears that life table \refs{mi-dlt-stare} is enough
to describe probabilistic structure for the model
$(\mathcal{S}^{\ast},\mathcal{T}^{\ast})$
presented in Figure~\ref{Fig.2}b
even though this table does not contain
$l_{x+k}^{7}$, $d_{x+k}^{78}$ and $l_{x+k}^{9}$, $d_{x+k}^{9 10}$.
This is possible due to the fact that
both states $7$ and $9$ are reflex (i.e. $i$ is
transient and $q^{\ast}_{ii}(k)=0$) for which there exists only
one possibility to leave. Hence $l_{x+k}^{7}$ and
$d_{x+k}^{78}$ are unambiguously defined by $d_{x+k}^{17}$ and
$d_{x+k}^{27}$ in the following way
$l_{x+k}^{7}=d_{x+k-1}^{17}+d_{x+k-1}^{27}=d_{x+k}^{78}.$
Correspondingly $l_{x+k}^{9}$ and $d_{x+k}^{9 10}$ are connected
by the relation
$l_{x+k}^{9}=d_{x+k-1}^{39}+d_{x+k}^{49}+d_{x+k-1}^{59}+d_{x+k}^{69}=d_{x+k}^{9
10}.$

The transition matrix of $\{X^{\ast}(k)\}$ for DD disease insurance
model presented in Figure~\ref{Fig.2}b has the following form
\begin{eqnarray}\label{macierzQ}
\bfQ^{\ast}(k)= \left(
 \begin{array}{cccccccccc}
  q^{\ast}_{11}(k) & q^{\ast}_{12}(k) & q^{\ast}_{13}(k) &                0 &                0 &                0 & q^{\ast}_{17}(k) & 0 &                0 & 0 \\
                 0 & q^{\ast}_{22}(k) & q^{\ast}_{23}(k) &                0 &                0 &                0 & q^{\ast}_{27}(k) & 0 &                0 & 0 \\
                 0 &                0 &                0 & q^{\ast}_{34}(k) &                0 &                0 &                0 & 0 & q^{\ast}_{39}(k) & 0 \\
                 0 &                0 &                0 &                0 & q^{\ast}_{45}(k) &                0 &                0 & 0 & q^{\ast}_{49}(k) & 0 \\
                 0 &                0 &                0 &                0 &                0 & q^{\ast}_{56}(k) &                0 & 0 & q^{\ast}_{59}(k) & 0 \\
                 0 &                0 &                0 &                0 &                0 &                0 &                0 & 0 &                1 & 0 \\
                 0 &                0 &                0 &                0 &                0 &                0 &                0 & 1 &                0 & 0 \\
                 0 &                0 &                0 &                0 &                0 &                0 &                0 & 0 &                0 & 1 \\
                 0 &                0 &                0 &                0 &                0 &                0 &                0 & 0 &                0 & 1
 \end{array}
 \right),
\end{eqnarray}
where
\begin{eqnarray}
q_{ij}(k)= \left\{
\begin{array}{lll}
\frac{l^{i}_{x+k+1}- \sum_{j:(i,j) \in \mathcal{T}} d^{ij}_{x+k}}{l^{i}_{x+k}} &{\rm for}& j=i  \nonumber \\
\frac{d^{ij}_{x+k}}{l^{i}_{x+k}} &{\rm for}& j \neq i  \nonumber
\end{array},
\right.
\end{eqnarray}
and $l_{x+k}^{i}$, $d_{x+k}^{ij}$ come from  \refs{mi-dlt}.

If a multiple state life table is not available, then estimation of
$\bfQ^{\ast}(k)$ is needed.
We refer to \cite{DZ2015} where this problem
is analysed in detail in case of lung cancer disease for
multiple increment-decrement tables \refs{mi-dlt-stare}.

\subsection{Net premiums}\label{sec:ap.n.p}

In what follows we analyse three scenarios, where
\begin{description}
\item[-] premium is paid only if $X^{\ast}(k)=1$ (the insured is healthy / active), i.e. ${\mathcal{S}_{p}}=\{1\}$,
\item[-] premium is paid only if $X^{\ast}(k)=1,2$ (the insured is healthy or has not a lung cancer with distant metastases), i.e. ${\mathcal{S}_{p}}=\{1,2\}$,
\item[-] premium is paid if $X^{\ast}(k)=1,2,\ldots, 6$ (the insured is alive, independently on his health status), i.e. ${\mathcal{S}_{p}}=\{1,2,3,4,5,6\}$.
\end{description}

Then, by Theorem~\ref{Tw1}, the net period
premium, paid during the first $m$ units of the insurance
contract, has the following form
\begin{eqnarray}\label{skladkap.DD}
p_{\mathcal{S}_p}= \left\{
\begin{array}{lcl}
\frac{\bfM^T Diag \left(\bfC_{in} \bfD^T \right)\bfS}{
\ddot{a}_{1(1)}(0,m-1)}              & \textrm{for}   & {\mathcal{S}_{p}}=\{1\} \ \ \textrm{and} \ \ 1 \leq  m \leq n  \\

\frac{\bfM^T Diag \left(\bfC_{in} \bfD^T \right)\bfS}{
\ddot{a}_{1(1)}(0,m-1)+\ddot{a}_{1(2)}(1,m-1)}         & \textrm{for} & {\mathcal{S}_{p}}=\{1,2 \}  \ \ \textrm{and} \ \  2 \leq  m \leq n  \\

\frac{\bfM^T Diag \left(\bfC_{in} \bfD^T \right)\bfS}{
\ddot{a}_{1(1)}(0,m-1)+\ddot{a}_{1(2)}(1,m-1)+
\sum_{i=3}^{6}\ddot{a}_{1(i)}(i-2,m-1)}                &
\textrm{for}& {\mathcal{S}_{p}}=\{1,...,6\}  \ \ \textrm{and} \ \
5 \leq  m \leq n
\end{array} \right. ,
\end{eqnarray}
where $\ddot{a}_{1(i)}(t,m-1)$ is defined  by \refs{a-1i}.

Premiums calculated in Section~\ref{ADBs_insurance}
and Section~\ref{Ad_insurance} are based on
formulas \refs{sk³.jednoraz.} and \refs{skladkap.DD}, where
\begin{description}
\item{$\bfM$} is described under the assumption, that the interest rate is constant and equal $ 1 \%$,
\item{$\bfD$} is calculated for a 40-year-old person ($x=40$) and a 25-year-insurance period ($n=25$), based on
results presented in Section~\ref{sec:lc-ex} (apart from Table~\ref{Tab.NP2}, where
matrix $\bfD$ is calculated for 20, 30, 50 and 60 age at entry additionally).
\end{description}
We assume that period premiums are paid up to the end
of insurance period ($n=m=25$).

\subsubsection{Accelerated benefit for temporary life insurances} \label{ADBs_insurance}

We assume that the living benefit provides an
acceleration of part $\lambda$ of the basic life cover $1$ unit.
If the insured person's death occurred in time interval $[k,k+1)$, $k=0,1,2,...,n-1$, before the end of the insurance contract, then
at time $k+1$ the insurer pays benefit $1$ (i.e. $c=1$). Then cash flows matrix, which
consists only of an income to the total loss found, has the form
\begin{eqnarray*}
\bfC_{in}= \left(
 \begin{array}{cccccccccc}
0   &0   &0         &0   &0   &0   &0   &0  &0            &0  \\
0   &0   &\lambda   &0   &0   &0   &1   &0  &0            &0  \\
0   &0   &\lambda   &0   &0   &0   &1   &0  &1 - \lambda  &0  \\
\multicolumn{10}{c}{\dotfill}\\
0   &0   &\lambda   &0   &0   &0   &1   &0  &1 - \lambda  &0
 \end{array}
\right) \in \RL^{26 \times 10}.
\end{eqnarray*}

The premiums for such an insurance, depending on acceleration
parameter $\lambda$ and sex of the insured person, are presented
in Table~\ref{Tab.NP1}.
\begin{table}
 \begin{center}
  \caption{Premiums for acceleration benefit for temporary life insurances.}
\begin{tabular}{|l|c|c|c|c|c|c|c|c|} \hline
Premium&\multicolumn{2}{|c|}{$\pi$}&\multicolumn{2}{|c|}{$p_{
\{1\}}$}&\multicolumn{2}{|c|}{$p_{
\{1,2\}}$}&\multicolumn{2}{|c|}{$p_{ \{1,2,3,4,5,6\}}$} \\ \hline
\label{Tab.NP1} $\lambda$  &Woman   &Man     &Woman   &Man
&Woman   &Man     &Woman   &Man      \\ \hline 0.001
&0.10616 &0.22030 &0.00497 &0.01085 &0.00496 &0.01083 &0.00496
&0.01082 \\ \hline 0.25        &0.10638 &0.22083 &0.00498 &0.01087
&0.00498 &0.01086 &0.00497 &0.01084 \\ \hline 0.5         &0.10660
&0.22137 &0.00499 &0.01090 &0.00499 &0.01088 &0.00498 &0.01087 \\
\hline 0.75        &0.10682 &0.22190 &0.00500 &0.01092 &0.00500
&0.01091 &0.00499 &0.01090 \\ \hline 0.9…99      &0.10704 &0.22243
&0.00501 &0.01095 &0.00501 &0.01093 &0.00500 &0.01092 \\ \hline 1
&0.10704 &0.22243 &0.00501 &0.01095 &0.00501 &0.01093 & ---    &
---    \\ \hline
\end{tabular}
 \end{center}
\end{table}
Note that the premiums for women are about $51.8\%$ lower than for
men. It appears that the acceleration parameter does not have a big impact on the premium,
for example premiums for $\lambda=0,001$ and $\lambda=1$ differ
relatively by less than $1 \%$, both for women and men.
Analogously the relative difference between $p_{\{1\}}$ and $p_{\{1,2,3,4,5,6\}}$
is about $0.15 \%$ for women and $0.26 \%$ for men.

Net single premiums depending on the age of insured person are presented in Table~\ref{Tab.NP2}.
\begin{table}
 \begin{center}
  \caption{Net single premiums depending on the age at entry($\lambda=0.5, n=25$)}
\begin{tabular}{|l|c|c|c|} \hline
$x$ &$\pi^{woman}$  & $\pi^{man}$   & $(\frac{\pi^{man}}{\pi^{woman}}-1) \cdot 100$ \\ \hline \label{Tab.NP2}
20  &0.01602 &0.04452 & 178  \\ \hline
30  &0.04468 &0.10437 & 134  \\ \hline
40  &0.10660 &0.22137 & 108  \\ \hline
50  &0.23262 &0.39691 & 71   \\ \hline
60  &0.52662 &0.63972 & 21   \\ \hline
\end{tabular}
 \end{center}
\end{table}
It is not surprising that the values of premiums for men and women are increasing with the age at entry,
but the calculations show a strong influence of sex on the amount of net premiums.
We can observe that regardless of age net single premiums are lower for women.
The difference between premiums for young male and female decreases with the rise of the age at entry up to
$178\%$.

\subsubsection{Additional DD benefit for life insurances}\label{Ad_insurance}

Let $c^{ad}(k)$ denote the lump sum benefit payable at time $k$
on condition that the insured person's irreversible phase of dread
disease occurred in time interval $[k-1,k)$, $k=1,2,...,n-1$ and
$X^{\ast}(k)=3$, enabling the use of a more expensive and more
complete diagnosis. Sometimes, the single cash payment $c^{ad}(k)$ is replaced by a
series of payments  $b_j(k)$ (the annuity payable for period
$[k,k+1)$ if the insured is terminally ill at time $k$, i.e. $X^{\ast}(k)=3,4,5,6$)
conditional on the survival of the insured.

Moreover, let $c(k+1)$ denote the benefit payable at time $k+1$
if the insured person's death occurred in time interval $[k,k+1)$,
$k=0,1,2,...,n-1$ and $X^{\ast}(k)=7,9$, before the end of the
insurance contract.

Let $d$ be the pure endowment benefit payable at time $n$ if the
insured person is still alive at that time (i.e. $X^{\ast}(n)=1,2,3,4,5,6$).

We combine DD insurance for $\lambda=0$ with life insurance and analyse
the following cases:
\begin{description}
\item[Case 1] Additional lump sum DD benefit for $n$-year temporary life insurance \\
              ($c^{ad}(k)=1$ and $c(k+1)=1$)
\begin{eqnarray*}
\bfC_{in}= \left(
 \begin{array}{cccccccccc}
0   &0   &0   &0   &0   &0   &0   &0  &0  &0  \\
0   &0   &1   &0   &0   &0   &1   &0  &0  &0  \\
0   &0   &1   &0   &0   &0   &1   &0  &1  &0  \\
\multicolumn{10}{c}{\dotfill}\\
0   &0   &1   &0   &0   &0   &1   &0  &1  &0
 \end{array}
\right) \in \RL^{26 \times 10}.
\end{eqnarray*}
\item[Case 2] Additional annuity DD benefit for $n$-year temporary life insurance \\
              ($b_j(k)=0.25$ for $k=n, n-1, ..., j-2$, $j=3,4,5,6$ and $c(k+1)=1$)
\begin{eqnarray*}
\bfC_{in}= \left(
 \begin{array}{cccccccccc}
0   &0   &0      &0      &0      &0      &0   &0  &0  &0  \\
0   &0   &0.25   &0      &0      &0      &1   &0  &0  &0  \\
0   &0   &0.25   &0.25   &0      &0      &1   &0  &1  &0  \\
0   &0   &0.25   &0.25   &0.25   &0      &1   &0  &1  &0  \\
0   &0   &0.25   &0.25   &0.25   &0.25   &1   &0  &1  &0  \\
\multicolumn{10}{c}{\dotfill}\\
0   &0   &0.25   &0.25   &0.25   &0.25   &1   &0  &1  &0
 \end{array}
\right) \in \RL^{26 \times 10}.
\end{eqnarray*}
\item[Case 3] Additional lump sum DD benefit for $n$-year endowment insurance \\
              ($c^{ad}(k)=1$, $c(k+1)=1$, $d=1$)
\begin{eqnarray*}
\bfC_{in}= \left(
 \begin{array}{cccccccccc}
0   &0   &0   &0   &0   &0   &0   &0  &0  &0  \\
0   &0   &1   &0   &0   &0   &1   &0  &0  &0  \\
0   &0   &1   &0   &0   &0   &1   &0  &1  &0  \\
\multicolumn{10}{c}{\dotfill}\\
0   &0   &1   &0   &0   &0   &1   &0  &1  &0  \\
1   &1   &2   &1   &1   &1   &1   &0  &1  &0
 \end{array}
\right) \in \RL^{26 \times 10}.
\end{eqnarray*}

\end{description}

The premiums for the described cases, depending on sex of the insured
person, are presented in Table~\ref{Tab.NP3}.
\begin{table}
 \begin{center}
  \caption{Premiums for combination additional DD benefit and life insurances.}
\begin{tabular}{|l|c|c|c|c|c|c|c|c|} \hline
Premium&\multicolumn{2}{|c|}{$\pi$}&\multicolumn{2}{|c|}{$p_{
\{1\}}$}&\multicolumn{2}{|c|}{$p_{
\{1,2\}}$}&\multicolumn{2}{|c|}{$p_{ \{1,2,3,4,5,6\}}$} \\ \hline
\label{Tab.NP3} Sex    &Woman       &Man         &Woman       &Man
&Woman       &Man         &Woman       &Man      \\ \hline Case 1
&0.11610    &0.23865    &0.00544    &0.01175    &0.00543
&0.01173    &0.00543    &0.01172 \\ \hline Case 2 &0.10918
&0.22599    &0.00511    &0.01113    &0.00511    &0.01111
&0.00510    &0.01110 \\ \hline Case 3 &0.79813    &0.81671
&0.03736    &0.04021    &0.03733    &0.04015    &0.03731
&0.04010  \\  \hline
\end{tabular}
 \end{center}
\end{table}
Note that replacement of the single cash payment $c^{ad}_{k}$ (Case 1)
with a series of payments  $b_k$ (Case 2) relatively reduces
premiums by about $6\%$ for both females and for males.

\section{Discussion}
\label{sec:conc}

Combination of life insurance contract and
supplementary insurances leads to complex protecting packages.
The $({\mathcal{S}}^{\ast},{\mathcal{T}}^{\ast})$ model considered in this paper
also allows to incorporate options related to the health status of the
insured, such as disability or permanent inability to work.
Additionally, it is possible to use the part of the extended multiple state model
concerning the population of those suffering from lung cancer with metastasis (e.g. states $3, 4, 5, 6, 9,10$)
to derive the value of viatical settlement payments under the condition that the insured person is at state $3$.

The matrix approach to valuation of insurance contracts can be used
for other insurance contracts which are modelled by the multiple state model.
In particular, it can be useful for a {\it general buy-back accelerated critical illness} model presented in \cite{Bri10}.
Moreover, using the same ideas as in the proof of Theorem 1 combined with the Dijkstra's algorithm, one can determine the value
of annuity paid in any subset of states belonging to the state space ${\mathcal{S}}^{\ast}$
(for marriage insurance or marriage revers annuity contracts).

Numerical analysis of insurance risk for lung cancer shows a significant impact of gender for the calculation of premium.
Due to the high mortality rate of people suffering from lung cancer disease,
there is very little cost difference between contracts paid according to the scenario
when premiums are paid only by a healthy insured and the scenario when premiums are paid independently
of the health status of the insured (just by a living person).


\vspace{1cm}

\begin{thebibliography}{99}

\bibitem{Bri10}
Brink, A. (2010). Practical example of split benefit accelerated Critical Illness insurance product.
In {\it Transitions of the 29th International Congress of Actuaries} Cape Town, South Africa, p.677-703.

\bibitem{DG93}
Dash A., Grimshaw D. (1993) Dread Disease cover - an actuarial
perspective. {\it Journal of the Staple Inn Actuarial Society},
Vol. 33, p. 149-193.

\bibitem{NHF}
Data base of histories of hospitalization from Lower Silesia
Department of National Health Fund - unpublished due to
confidentially.

\bibitem{Deb03}
D\c{e}bicka, J. (2003) Moments of the cash value of future payment
streams arising from life insurance contracts, {\it Insurance:
Mathematics and Economics}, {\bf 33}, p. 533-550.

\bibitem{Deb13}
D\c{e}bicka J. (2013) An approach to the study of multistate
insurance contracts. {\it Applied Stochastic Models in Bussines
and Industry} Vol. 29, Issue 3, p. 224-240.

\bibitem{DZ2015}
D\c{e}bicka J., Zmy\'{s}lona, B. Modelling of lung cancer
survival data for critical illness insurances.
Manuscript available at arXiv:1602.08696.


\bibitem{Dij59}
Dijkstra E. W. (1959) A note on two problems in connexion with
graphs. {\it Numerische Mathematik}, {\bf  1}, p. 269–271

\bibitem{Fre90}
Frees E.W. (1990) Stochastic life contingencies with solvency
considerations. {\it Transactions of the Society of Actuaries},
{\bf XLII }, p. 91-148.

\bibitem{HP99}
Haberman S., Pitacco E. (1999) {\it Actuarial Models for
Disability Insurance}, Chapman \& Hall/CRC.

\bibitem{Hoe69}
Hoem J.M. (1969) Markov Chain Models in Life Insurance, {\it
Bl\"{a}tter der Deutschen Gesellschaft f\"{u}r
Versicherungsmathematik}, Vol. IX, p. 91-107.

\bibitem{Hoe88}
Hoem J.M. (1988) The Versality of the Markov Chain as a Tool in
the Mathematics of Life Insurance, {\it Transactions of the 23rd
International Congress of Actuaries}, Helsinki, Vol. R, p.171-202.

\bibitem{Par94}
Parker, G. (1994) Stochastic Analysis of Portfolio of Endowment
Insurance Policies, {\it Scandinavian Actuarial Journal}, {\bf 77}
(2), p. 119-130.

\bibitem{Pit94}
Pittaco E. (1994) LTC insurance. From the multistate model to
practical implementations. {\it Proceedings of the XXV ASTIN
Colloquium}, Cannes, Frances. p. 437-452.

\bibitem{Pit14}
Pitacco E. (2014) {\it Health Insurance. Basic Actuarial
Models.}, EAA Series, Springer.

\bibitem{TTZ-GUS} {Life Tables of Poland 2008}. Available at
www.stat.gov.pl/en/topics/population/life-expectancy/

\bibitem{Wat84}
Waters H.R. (1984) An approach to the study of multiple state
models, {\it Journal Institute of Actuaries}, Vol. 111, Part II,
No. 448, p. 363-374.

\bibitem{WD14}
Wojciechowska U., Didkowska J. (2014) Zachorowania i zgony na
nowotwory z³oœliwe w Polsce. Krajowy Rejestr Nowotworów, Centrum
Onkologii - Instytut im. Marii Sk³odowskiej - Curie. Available at:
www.onkologia.org.pl/raporty/ (date 10.04.2014).


\bibitem{Wol94}
Wolthuis H. (1994) {\it Life insurance mathematics (The Markovian
model)}, CAIRE Education Series, No. 2, Bruxelles.


\end{thebibliography}
\end{document}